\newcommand{\fg}[1]{Fig.\,\ref{fig:#1}}

\newcommand{\dd}{\mathrm{d}}
\newcommand{\lna}{\ln\!A}

\documentclass{PoS}
\usepackage{siunitx}

\newlength{\bibitemsep}
\newlength{\bibparskip}
\let\oldthebibliography\thebibliography
\renewcommand\thebibliography[1]{%
  \oldthebibliography{#1}%
  \setlength{\parskip}{\bibitemsep}%
  \setlength{\itemsep}{\bibparskip}%
  \footnotesize%
}

\title{Data-driven model of the cosmic-ray flux and mass composition from 10\,GeV to $10^{11}$\,GeV}

\ShortTitle{Data-driven model of the cosmic-ray flux and composition}

\author{\speaker{Hans Peter Dembinski}\\
        Max Planck Institute for Nuclear Physics, Heidelberg, Germany\\
        E-mail: \email{hdembins@mpi-hd.mpg.de}}

\author{Ralph Engel\\
        Karlsruhe Institute of Technology, Karlsruhe, Germany\\
        E-mail: \email{ralph.engel@kit.edu}}

\author{Anatoli Fedynitch\\
        DESY Zeuthen, Zeuthen, Germany\\
        E-mail: \email{anatoli.fedynitch@desy.de}}

\author{Thomas Gaisser\\
        Bartol Institute, University of Delaware, Newark DE, USA\\
        E-mail: \email{gaisser@bartol.udel.edu}}

\author{Felix Riehn\\
        LIP, Lisbon, Portugal\\
        E-mail: \email{friehn@lip.pt}}

\author{Todor Stanev\\
        Bartol Institute, University of Delaware, Newark DE, USA\\
        E-mail: \email{stanev@bartol.udel.edu}}

\abstract{We present a new parametrization of the cosmic-ray flux and its mass composition over an energy range from 10\,GeV to $10^{11}$\,GeV. Our approach is data-driven and relies on theoretical assumptions as little as possible. We combine measurements of the flux of individual elements from high-precision satellites and balloon experiments with indirect measurements of mass groups from the leading air shower experiments. To our knowledge, we provide the first fit of this kind that consistently takes both statistical and systematic uncertainties into account. The uncertainty on the energy scales of individual experiments is handled explicitly in our mathematical approach. Part of our results is a common energy scale and adjustment factors for the energy scales of the participating experiments. Our fit has a reduced $\chi^2$-value of 0.5, showing that experimental data are in good agreement, if systematic uncertainties are considered. Our model may serve as a world-average of the measured fluxes for individual elements from proton to iron from 10\,GeV to $10^{11}$\,GeV. It is useful as an input for simulations or theoretical computations based on cosmic rays. The experimental uncertainties of the input data are captured in a covariance matrix, which can be propagated into derived quantities.}

\FullConference{35th International Cosmic Ray Conference -- ICRC2017\\
		10-20 July, 2017\\
		Bexco, Busan, Korea}

\begin{document}

\section{Introduction and motivation}


The cosmic-ray flux spans more than 11 decades in energy. Individual experiments cover only a part of that range, which means that a common picture needs to be pieced together from many data sets. The {\em Database of Charged Cosmic Rays} (CRDB)~\cite{Maurin:2013lwa} is a commendable effort in providing central machine-readable access to cosmic-ray data from satellite and balloon experiments. These experiments are able to separate elements by their charge $Z$ and report the flux per element.

Above a few 100\,TeV, direct observation is not feasible and ground-based experiments take over. These measure a cosmic ray indirectly through secondary particles produced in a cosmic-ray induced air shower. Air-shower experiments achieve large apertures, but loose the ability to discriminate the charge $Z$ of each cosmic ray. The distinguishing feature is now the mass $A$, which has a measurable effect on several air-shower observables, but stochastic fluctuations in the shower development overshadow these signatures. Ground-based experiments report the all-particle flux, and potentially the flux fractions contributed by elements which fall into a given mass range.

The inference of the cosmic-ray energy $E$ and mass $A$ from air-shower observables depends on theoretical models. The air-shower development is driven by soft QCD interactions, which are computed from phenomenological models tuned to collider data. The theoretical uncertainties in these models cause larger systematic uncertainties on the reported results compared to direct measurements. In particular, the energy scale of each experiment has a non-negligible uncertainty of about $(10-20)\,\%$, which produces apparent discrepancies between independent experiments.

We present a new data-driven model, called the \emph{Global Spline Fit} (GSF). It parametrizes the latest and most detailed measurements of the cosmic-ray flux and its composition from 10\,GeV to $10^{11}$\,GeV, combining direct and air-shower observations. Smooth curves of the fluxes of all elements from proton to nickel are provided. Energy-scales of included experiments are cross-calibrated based on the successive overlap of data sets in energy, and a common energy scale is established which is fixed by direct measurements. The GSF yields a covariance matrix which represents the experimental uncertainty of the input data. The covariance matrix can be used to compute standard deviations for quantities derived from the cosmic ray flux and composition.

In contrast to other recent models~\cite{Hoerandel_polygonato,GST_main_paper,Gaisser:2012em}, the flux in this work is not constructed from components with power-law shape and rigidity-dependent cut-offs. The GSF only assumes that the flux is smoothly varying and otherwise relies only on the available data. It is not a model in the usual sense since it does not try to explain the data, it just describes an average. This approach has become feasible recently, since air-shower experiments now cover the energy range above 1\,PeV with detailed data about the cosmic-ray mass composition.

The GSF model, as presented here, is used in a state-of-the-art calculation of atmospheric lepton fluxes also presented at this conference~\cite{Anatoli_lepton_flux}. In the following, we will describe the GSF approach, and discuss our preliminary results. The final data tables and code for the GSF model will be published later this year, allowing everyone to reproduce these results.

\section{Global Spline Fit}

\begin{figure}
    \centering
    \includegraphics[width=0.49\textwidth]{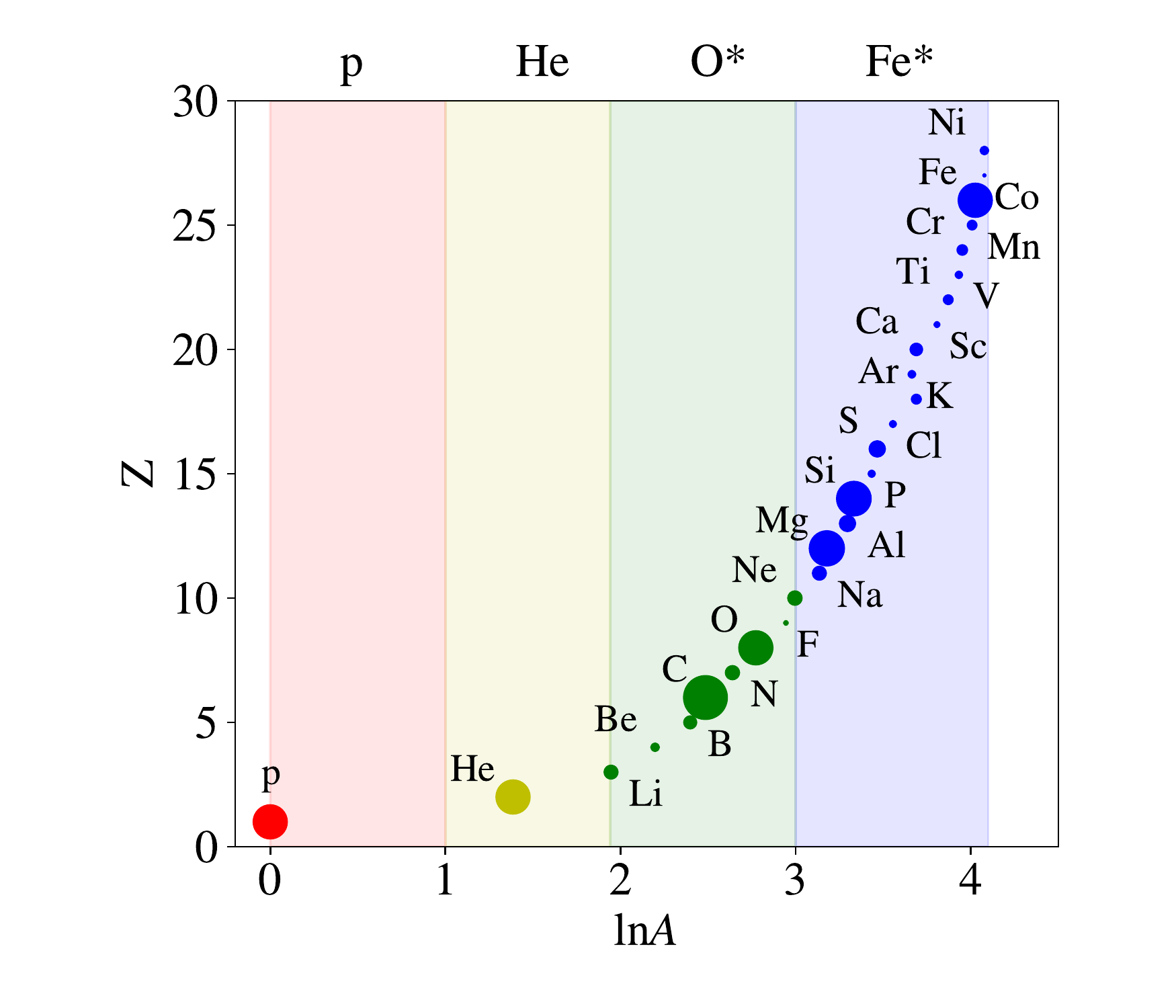}
    \hfill
    \includegraphics[width=0.49\textwidth]{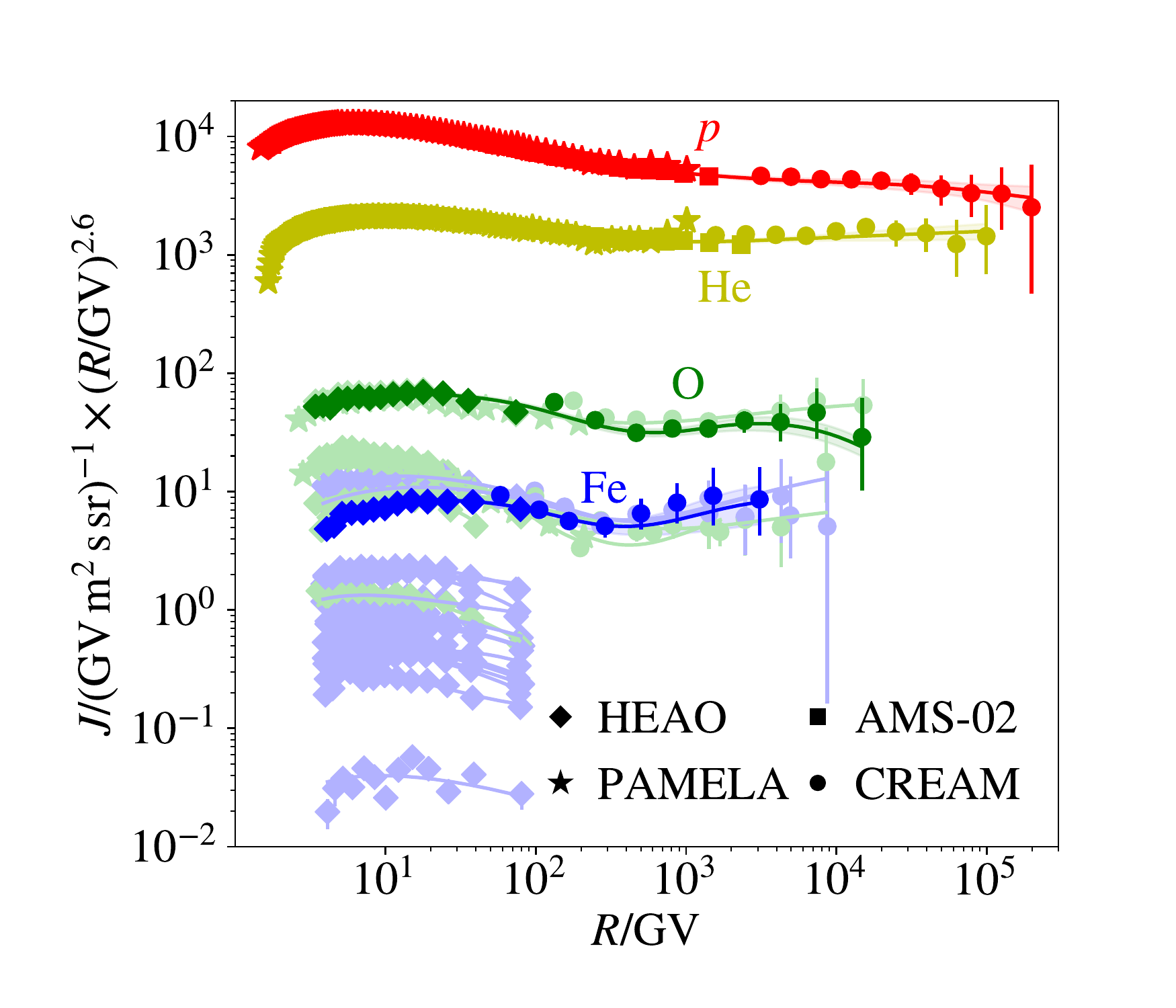}
    \caption{{\em Left:} Elements from proton to nickel are sorted into four mass groups, covering roughly equal intervals in logarithmic mass $\lna$. The groups are named after the leading element, which provides most to the differential flux $\propto \dd N/\dd E$ of the group. The size of the marker indicates the flux ratio $f_{iL}$ relative to the leading element $L$ of the group, as described in the text. The values are obtained from fits to HEAO data~\cite{HEAO_data}. {\em Right:} Fluxes of individual elements measured by satellites and balloons~\cite{HEAO_data,AMS02_p_data,AMS02_He_data,PAMELA_He_data,PAMELA_p_data,CREAM_heavy_data,CREAM_light_data}, together with fitted spline curves. Highlighted points and curves show the four leading elements, fainter points and curves sub-leading elements from the oxygen and iron groups.}\label{fig:mass_groups}
\end{figure}


In the GSF model, the cosmic-ray flux is divided into four mass groups, which cover roughly equal ranges in logarithmic mass $\lna$, as shown on the left-hand side in \fg{mass_groups}. We split in $\lna$, since air-shower measurements are sensitive to changes in $\lna$ rather than $A$. Each group has a leading element $L$ that contributes most of the flux per energy interval. We note that if two elements have the same abundance in flux per rigidity interval $J(R) \propto \dd N / \dd R$, the element with the higher charge contributes more to the flux per energy interval $\dd N / \dd E$. The leading elements are thus the heaviest abundant elements in each group; namely proton, helium, oxygen, and iron. The oxygen and iron groups contain many sub-leading elements. In the oxygen group, carbon contributes nearly as much as oxygen. In the GSF model, the flux $J_i(R)$ of a sub-leading element $i$ is kept in a constant ratio $f_{iL}$ to the leading element $L \in \{p, \mathrm{He}, \mathrm{O}, \mathrm{Fe}\}$ of its group, $J_i(R) = f_{iL} \times J_L(R)$.

This treatment is motivated empirically by low energy data, shown on the right-hand-side in \fg{mass_groups}. If the differential flux is plotted as a function of rigidity, the fluxes of neighboring elements roughly have the same shape. In other words, the flux ratio of elements within a group is a very slowly varying function of $\ln R$. We approximate by keeping this ratio constant.

This approximation is rough, but only applied to sub-leading elements. Leading elements are fitted to experimental data and pull sub-leading elements with them. For many quantities computed from the GSF model, the exact ratios of sub-leading elements are not important; for example, for the nucleon flux and mean-logarithmic mass. Sub-leading elements however cannot be neglected. The fluxes of sub-leading elements contribute roughly a factor of two enhancement over the elemental fluxes of oxygen and iron for the respective mass groups. Air-shower measurements can only distinguish mass groups, so this additional flux has to be included. Neglecting this leads to artifacts in the description of the transition from direct to air-shower measurements.

The differential flux $J_L(R)$ of each leading element $L$ is parametrized by a smooth curve. We use a modified spline curve, build from a linear combination of B-splines~\cite{DeBoor_splines} shaped by a power-law term,
\begin{equation}
J_L(R) = \sum_k a_{Lk} \times b_k \big(\ln (R/\si{GV})\big) \times (R/\si{GV})^{-3},
\label{eq:leading_spline}
\end{equation}
where $b_k \big(\ln (R/\si{GV})\big)$ are standard cubic B-splines defined over a common vector of knots, $a_{Lk}$ are the coefficients adjusted in a global fit to all data sets, and $(R/\si{GV})^{-3}$ is the shaping factor. The latter stabilizes numerical results. Without the shaping factor, the coefficients $a_{Lk}$ would vary over many orders of magnitude. This would lead to precision loss in numerical computations.

The knot locations form a regular grid in $\ln(R/\mathrm{GV})$, with a typical step size of $0.5$. We double (half) the step size in regions where the flux varies more (less). We tested that the exact placement of knots has negligible impact on our results by varying all knot positions by $\pm 0.1$.

Our model parametrizes the differential flux of nuclei per rigidity interval $J(R)$. Air-shower measurements of the cosmic-ray flux are reported as the differential flux of nuclei per energy interval $J(E)$. The latter is computed from the former as $J(E) = J(R)\, \dd R/\dd E$. The relationship between total energy $E$ and rigidity $R$ depends on the number of nucleons $A$ and protons $Z$ and is computed individually for each element. Air-shower measurements describe the flux of mass groups. We sum the flux of all elements in each group when comparing the model to such measurements.

We consider the effect of a potential energy-scale offset $z_E = (\tilde E-E)/E$ for each experiment, where $\tilde E$ is the energy reported by an experiment for the differential flux $\tilde J(\tilde E)$ reported by an experiment. If $J(E)$ is the true flux at the true energy $E$, the apparent flux is
\begin{equation}
\tilde J(\tilde E) = J(E) \frac{\dd E}{\dd \tilde E}
= J\left(\frac{\tilde E}{1+z_E}\right) \frac{1}{1 + z_E}.
\label{eq:cr_flux_energy_rescaled}
\end{equation}
For a power-law spectrum $J(E) \propto E^{-\alpha}$, the apparent flux is changed by a factor of $(1+z_E)^{\alpha-1}$ when plotted on the same graph. For example, a small energy-scale offset of 10\,\% results in an apparent increase of 18\,\% for a spectral index $\alpha = 2.7$. This effect is well known~\cite{Hoerandel_polygonato,GST_main_paper}, and has been corrected by hand before. To our knowledge, the GSF approach is the first that derives the energy-scale offsets from a global fit, so that the corrections are a result instead of an input.

With these ingredients, we can compute residuals between measurements and the GSF model, and find spline amplitudes which minimize these residuals. We use a global least-squares fit to all data sets at once. We numerically find the spline amplitudes $a_{Lk}$ and energy-scale offsets $z_{E,l}$ by minimizing the function
\begin{equation}
    F' = \sum_i F_i + \sum_i \left(\frac{z_{E,i}}{\sigma_{E,i}}\right)^2,
\end{equation}
where the term $F_i$ represents the sum of squared residuals between data points and the GSF model for an experiment $i$, which depend on the spline amplitudes and the energy-scale offset $z_{E,i}$ for that experiment. To compute residuals, we take both statistical and systematic uncertainties into account. We also include correlations of systematic uncertainties, because systematic effects typically affect several data points similarly. A detailed treatment of systematic uncertainties is very important in a fit of cosmic-ray data, since they are often the largest uncertainties.

Each offset $z_{E,i}$ is handled as a squared residual over the relative energy-scale uncertainty $\sigma_{E,i}$ of experiment $i$ and added to $F'$. If the offsets $z_{E,i}$ are zero (no energy-scale adjustment), the second sum is minimal, but the first sum may increase. The global fit finds a balance, where energy-scale offsets minimize the deviations of experiments from the model (and therefore from each other), but do not become too large compared to the energy-scale uncertainties. This approach is a consistent extension of the standard least-squares method~\cite{Barlow:2017xlo}.

\section{Data sets selected as input}

Many experiments provide cosmic-ray data~\cite{Kampert_cr_review}. For the GSF, we use only the most recent measurements which divide the flux into separate contributions from mass groups. In case of direct measurements, we favor satellite over balloon experiments, since satellites do not have to correct for interactions in the residual atmosphere above the balloon. We make exceptions from these rules only to fill gaps in our targeted energy range.

We use direct measurements from the HEAO satellite~\cite{HEAO_data}, PAMELA~\cite{PAMELA_He_data,PAMELA_p_data}, AMS-02~\cite{AMS02_p_data,AMS02_He_data}, CREAM-I and II~\cite{CREAM_heavy_data,CREAM_light_data}. Solar modulation is corrected based on the force-field approximation~\cite{Gleeson:1968zza}, using modulation parameters from the CRDB~\cite{Maurin:2013lwa, 2011JGRA..116.2104U}. We use air-shower measurements from ARGO-YBJ~\cite{ARGO_YBJ_pHe_data}, TUNKA~\cite{TUNKA_data,TUNKA_quest_ecalib}, IceCube~\cite{IceCube_data}, KASCADE-Grande (KG)~\cite{KASCADE_Grande_energy_scale,KASCADE_Grande_data}, Telescope Array (TA)~\cite{TA_data}, and the Pierre Auger Observatory (Auger)~\cite{Auger_flux_data,Auger_fraction_data}. At the time of this writing, Telescope Array has not published a split of their all-particle flux into individual mass groups. It does not pass our criteria, but is included for its special significance of providing an all-particle flux that covers six decades in energy. For ARGO-YBJ, we use the reported proton+helium flux only below 1\,PeV; above 1\,PeV the flux drops sharply, in contradiction to three other data sets.

For a global fit, it is important that experiments report systematic uncertainties. The composition results from TUNKA, IceCube, and KASCADE-Grande do not yet have published systematic uncertainties. Fitting data sets with and without systematic uncertainties leads to inconsistencies. It puts the former at a disadvantage; data sets with systematic uncertainties have less pull on the model, although an evaluation of systematic uncertainties increases the reliability of results.

To prevent such inconsistencies, we assign 10\,\% uncertainty to composition measurements reported by TUNKA, and 15\,\% to those of IceCube. These estimates are based on the observed deviations of these results from each other. We assign 20\,\% to the KASCADE-Grande composition, a preliminary conservative estimate suggested by the authors. This is not a satisfactory solution, of course, but one we suggest until final results with systematic uncertainties are published.

Composition results from air-shower experiments are sensitive to the hadronic interaction model used to interpret the raw data, and models vary in our data sets. A single-model interpretation over the whole energy range is currently not possible with the available data. The Pierre Auger Observatory provides three composition results for three different hadronic interaction models~\cite{Auger_fraction_data}. We average these here, and treat the model variation as a systematic uncertainty.

\section{Results and discussion}

\begin{figure}[p]
    \includegraphics[width=\textwidth,trim={0.8cm 0.7cm 0.25cm 0}]{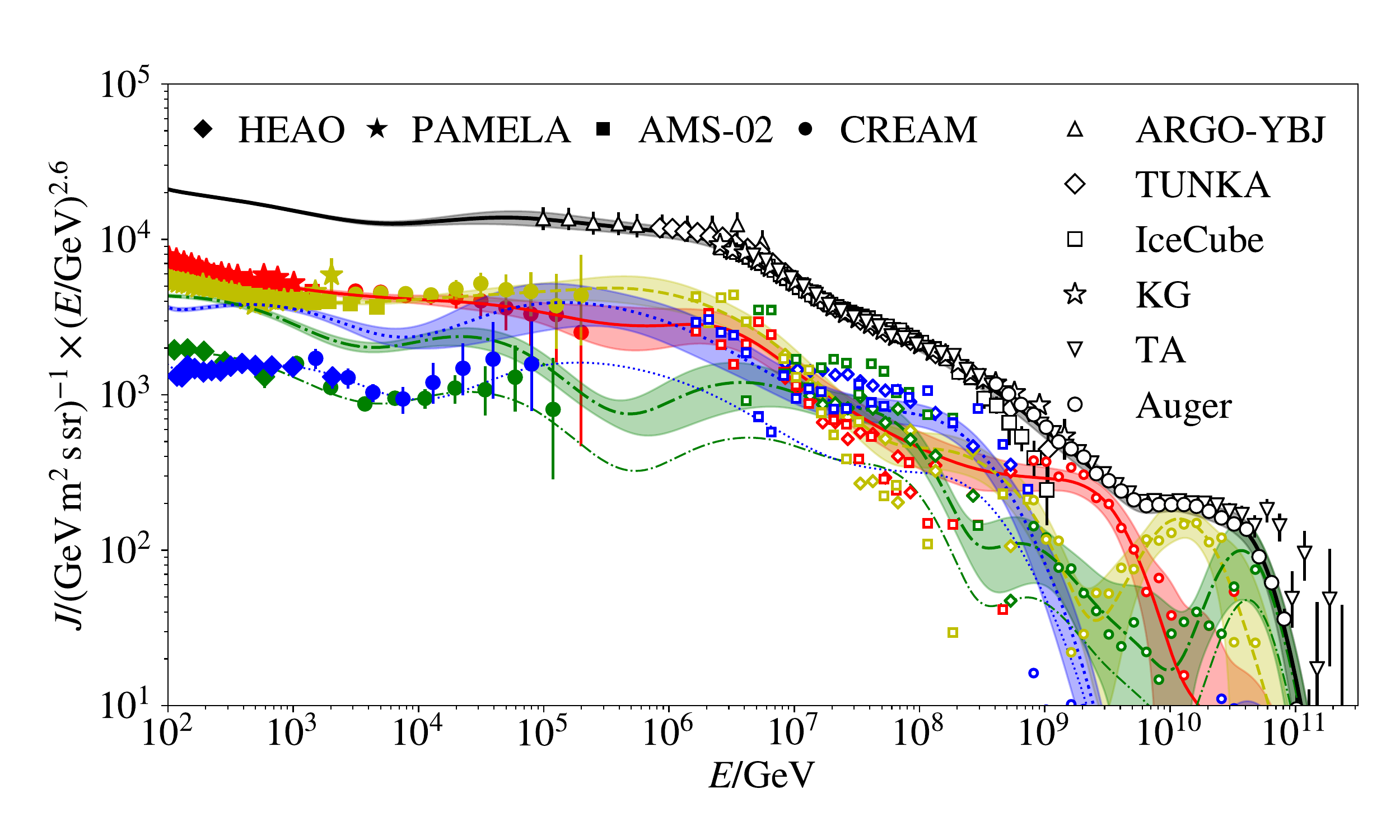}
    \caption{All-particle flux (black thick solid line), the flux contributed by protons (red line solid line), helium (yellow dashed line), the oxygen group (gray dash-dotted line), and the iron group (blue dotted line). Bands around the model lines show a variation of one standard deviation. Data points show measurements which were energy-scale adjusted as described in the text. Error bars represent combined statistical and systematic uncertainties. Data points of composition measurements from air-showers are not shown without error bars for clarity. In case of oxygen and iron, both the elemental flux and the group flux are shown; the smaller flux without error band is the elemental flux in each case. TA stands for Telescope Array, KG for KASCADE-Grande.}
    \label{fig:flux_overview}
\end{figure}

\begin{figure}[p]
    \includegraphics[width=0.59\textwidth]{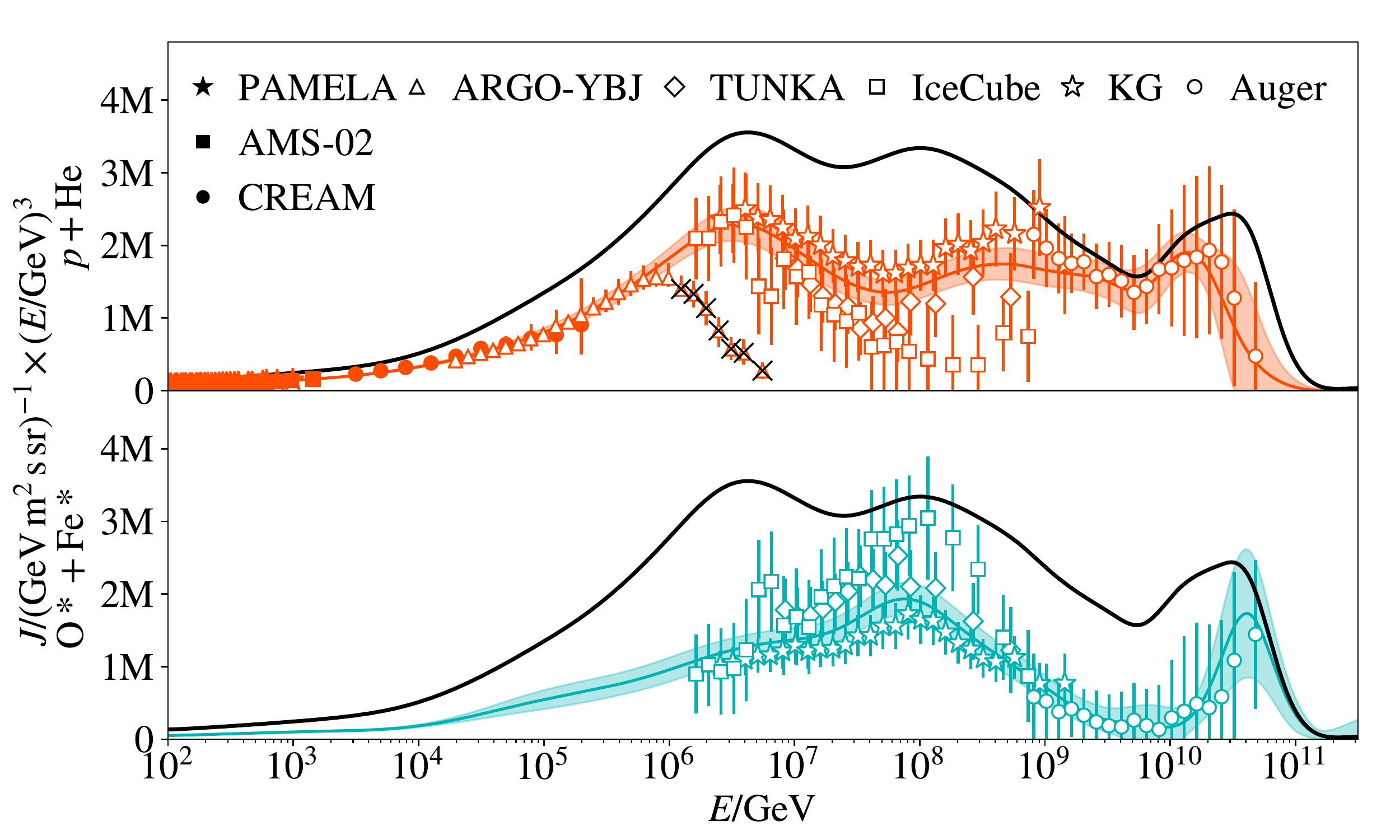}
    \includegraphics[width=0.4\textwidth]{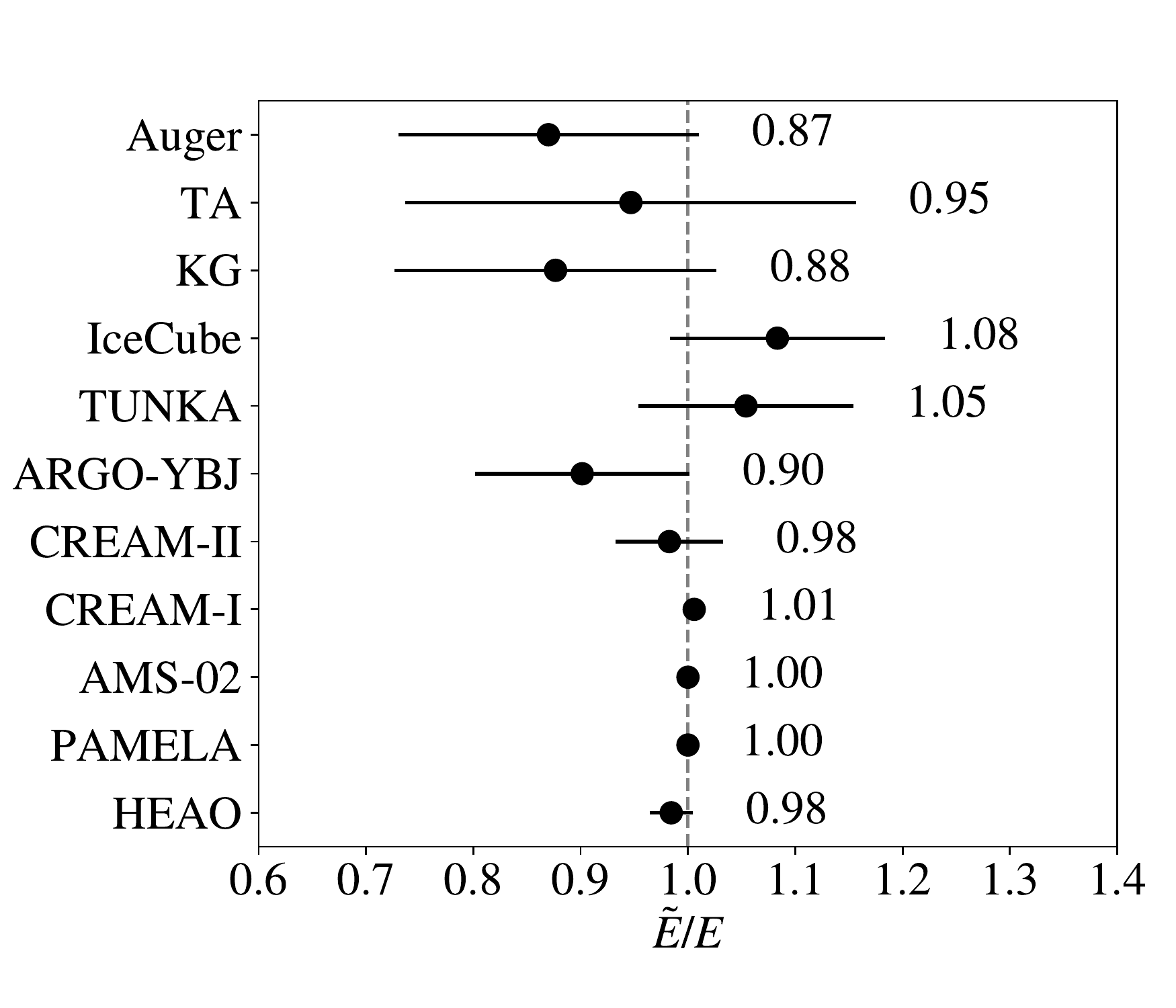}
    \caption{\emph{Left:} Particle flux in linear scale, split into a light (proton and helium) and heavy (others elements) component. The black solid line represents the all-particle flux. KASCADE-Grande and ARGO-YBJ reported their composition measurements in this split. For other experiments, synthetic points are generated for visual comparison from the more detailed composition data. The ARGO-YBJ points marked with crosses are not used in the fit. \emph{Right:} Ratios of energy scales used by an experiment relative to the cross-calibrated energy scale established by the GSF model. Error bars represent the reported systematic uncertainties of the energy scales. TA stands for Telescope Array, KG for KASCADE-Grande.}
    \label{fig:flux_split}
\end{figure}

The fitted GSF model is shown in \fg{flux_overview} and \fg{flux_split}. The global fit together with the adjusted energy-scales reveals detailed structure in the all-particle flux. We observe an unnamed dip around 5\,TeV, the \emph{knee} around 3\,PeV, a \emph{second knee} around 100\,PeV, the \emph{ankle} around 8\,EeV, a slight hardening around 20\,PeV, and finally the \emph{toe} around 60\,EeV. The energy-scale offsets found by the GSF are compatible with the reported systematic uncertainties.

We quantify the agreement of the GSF with the input data by computing the sum of squared residuals, taking both statistical and systematic uncertainties into account. We obtain a $\chi^2$ of 385.2 for 724 degrees of freedom, which indicates a good fit and implies that the data sets are overall consistent, if systematic uncertainties are taken into account. We remind, however, that this is achieved by rejecting part of the ARGO-YBJ data, and assigning $(10 - 20)\,\%$ systematic uncertainty to results where none were reported.

Telescope Array and the Pierre Auger Observatory show remarkable agreement, except for a notable discrepancy in the all-particle flux at the toe. It has been noted that these experiments observe different hemispheres and so the flux could be different, but so far there is no conclusive evidence for this hypothesis~\cite{Auger_flux_data}. The GSF model is build on the assumption that the flux is isotropic and therefore pulled towards the Auger data points, which have much smaller uncertainties.

Based on the GSF view on the flux and its mass composition, we can make a few observations. The first knee is created by a maximum in the proton and helium components, the second knee mostly by a maximum in the iron mass group. The mass composition data from the Pierre Auger Observatory at the high end of the cosmic-ray spectrum matches a Peters cycle~\cite{Peters1961} with a rigidity-dependent cut-off. The proton component drops before the ankle, so that the toe is dominated by helium and the oxygen group. The flux has a unique proton-maximum around 2\,EeV, before the ankle. We note that the all-particle flux between the second knee and the ankle looks like a simple power-law, although the mass composition changes drastically in the same energy range.

The GSF model is a framework to compute a world-average of the cosmic-ray flux and elemental composition, inspired by what the Particle Data Group does for various other measurements. We believe that a comprehensive model based on experimental data is a valuable input for many other analyses that compute derived quantities, like the atmospheric lepton flux. The GSF model summarizes the data provided by leading experiments, providing a combined estimate and a covariance matrix to compute the standard deviation of derived quantities. It establishes an common energy scale by cross-calibrating overlapping data sets, which is fixed by satellite and balloon experiments.

\section{Acknowledgments}

We acknowledge the valuable discussions with Andreas Haungs, Sven Schoo, Sergey Epimakhov, Serap Tilav, and Felipe Penha in preparation of this work. We further thank Javier G. Gonzales and Ioana Mari\c{s} for computer code that helped to kickstart this work. AF thanks for funding from the European Research Council (ERC) under the European Union's Horizon 2020 research and innovation programme (Grant No. 646623). TG and TS acknowledge support from the U.S. National Science Foundation (PHY-1505990).

\bibliographystyle{JHEP}
\bibliography{references}

\end{document}